\documentclass{article} 
\usepackage{iclr2021_conference,times}
\usepackage{natbib}
\usepackage{imakeidx}
\usepackage{amsmath}
\usepackage{caption,subcaption}
\usepackage{mathtools}
\usepackage{xfrac}
\usepackage{wrapfig}
\usepackage{booktabs}

\usepackage{amsmath,amsfonts,bm}

\def\eqref#1{equation~\ref{#1}}

\def\1{\bm{1}}

\DeclareMathAlphabet{\mathsfit}{\encodingdefault}{\sfdefault}{m}{sl}
\SetMathAlphabet{\mathsfit}{bold}{\encodingdefault}{\sfdefault}{bx}{n}

\usepackage{url}

\title{Improving Solar Cell Metallization Designs using Convolutional Neural Networks}

\author
{S. Bhattacharya$^{\dagger\ddagger\ast}$, D. Arya$^{\dagger\upsilon}$, D. Bhowmick$^{\dagger}$, R.~M. Thomas$^{\dagger\epsilon}$, D.~K. Gupta$^{\dagger\star\ast}$ \\
$^{\dagger}$Spectrum AI, Amsterdam, The Netherlands\\
$^{\ddagger}$Indian Institute of Technology (ISM) Dhanbad, India \\
$^{\upsilon}$Informatics Institute, University of Amsterdam, The Netherlands \\
$^{\epsilon}$Amsterdam UMC, University of Amsterdam, The Netherlands \\
$^{\star}$Transmute AI Research, Amsterdam, The Netherlands
}

\iclrfinalcopy 
\begin{document}

\maketitle

\begin{abstract}
Optimizing the design of solar cell metallizations is one of the ways to improve the performance of solar cells. Recently, it has been shown that Topology Optimization (TO) can be used to design complex metallization patterns for solar cells that lead to improved performance. TO generates unconventional design patterns that cannot be obtained with the traditional shape optimization methods.  

In this paper, we show that this design process can be improved further using a deep learning inspired strategy. We present SolarNet\footnote{Code is publicly available at \texttt{https://github.com/transmuteAI/SolarNet}.\\ $^{\ast}$Equal Contribution.}, a CNN-based reparameterization scheme that can be used to obtain improved metallization designs. SolarNet modifies the optimization domain such that rather than optimizing the electrode material distribution directly, the weights of a CNN model are optimized. The design generated by CNN is then evaluated using the physics equations, which in turn generates gradients for backpropagation. SolarNet is trained end-to-end involving backpropagation through the solar cell model as well as the CNN pipeline. Through application on solar cells of different shapes as well as different busbar geometries, we demonstrate that SolarNet improves the performance of solar cells compared to the traditional TO approach. 
\end{abstract}

\section{Introduction}

Solar energy has proven to be one of the potential sustainable energy sources, however, a limitation is imposed by our ability to convert it to electricity in a cost-effective way. Several different strategies are being explored to tackle this issue, and one among these is to improve the metallization pattern on these cells. Choice of metallization determines the distribution of voltage on the surface of the solar cell, thereby controlling the amount of current as well as power generated \citep{burgers1999design}. 

The aspect outlined above has been addressed in several past research works. \cite{Flat1979} put forward the use of multi-level grid metallization thus improving performance of the solar cells. Method presented by \cite{Antonini2009} helped in shifting from conventional to non-conventional metallization patterns. \cite{Wen2010} presented a descriptive study of different top contact grid structures. Most of the works in this domain have been restricted to preset designs, e.g. `H-pattern'. Recently, \cite{Gupta2015smo} presented the application of topology optimization methods (TO) capable of optimizing the front metallization designs with no apriori assumption on the shapes. Due to its flexibility, TO allows to design more efficient solar cells for any given variable set of environmental conditions, arbitrary choice of cell geometries and busbar locations \citep{Gupta2016_2, Gupta2016} and concentrated photovoltaics (CPV) with non-uniform illuminations \citep{Gupta_18}.

There have been recent research works aiming to improve TO output through combining it with methods from the field of deep learning. The simple underlying idea of these methods is to predict a new design using a deep learning model that has been trained on a wide variety of TO problems \citep{Rawat, Takahashi}. However, unlike the common tasks from the field of deep learning and computer vision, where problem varies within a limited set of possibilities (\emph{e.g.}, limited set of classes in a classification problems), the problems of TO are governed by physics with an infinite bandwidth of possible configurations. To this end, most works that have used deep learning for TO have only shown applications on simple problems, such as 2D compliance minimization \cite{Yu}, and the potential of data-driven methodologies on large and more complex TO problems is still unexplored. 

Solar cell metallization design is a complex nonlinear problem from optimization point of view \citep{Gupta2015smo}. Hence, we argue that a straight-forward data-driven approach is not suited for it. In this paper, we present \emph{SolarNet}, a deep learning inspired reparameterization strategy for improving metallization designs of solar cells. Our approach builds up on the recent work of \cite{Hoyer}
that exploited the power of popular deep learning method in TO without converting TO into a data-driven problem. Rather than optimizing the density parameter, as in the standard TO methodology, SolarNet optimizes the weights of a CNN which in turn produces the density field. Through application on solar cells of different shapes and different busbar designs, we show that SolarNet outperforms the traditional TO approach for the solar cell metallization design problem. 

\section{Mathematical Model}
\label{headings}

In this section, we briefly outline the mathematical formulation for the problem of current and voltage modeling on the front surface of the solar cell. Further, we also describe how the power output of the solar cell is modeled. For full details, see \cite{Gupta2015smo} and \cite{Gupta_18}. 

We consider here the case of a simple solar cell model as described in \cite{Gupta2015smo}. The output current generated by the solar cell is denoted by $I$, $R_{L}$ denotes the load resistance connected to the cell, and $R_{S}$ refers to the series resistance between the cell and the external load. $R_{SH}$ denotes shunt resistance that may come into play in the presence of cracks in the cell. For the sake of simplicity, we neglect the effects of $R_{SH}$ and $R_{S}$. 
The net current $I$ flowing through the cell can be represented as
\begin{equation} 
       I = I_{L} - I_{D}
\end{equation}
   where $I_{L}$ denotes photo-illumination current and $I_{D}$
   denotes the dark current in the absence of photo-illumination.
Based on the definition of $I_D$ as provided in \cite{green1982solar}, $I$ and $V$ can then be related as:
\begin{equation} 
        I = I_{L} - I_{0}(\text{exp}(qV/k_{B}T) - 1),
\end{equation}
     where $I_{0}$ is the reverse bias current, $q$ is the electric charge, $V$ is the voltage, $k_{B}$ is Boltzmann constant and $T$ is the temperature of the cell.
     
The current $I$ generated in the cell is an aggregate of the current generated on the entire surface of the cell. Thus, for accurate modeling on the front surface, the local phenomena needs to be considered. For this purpose, we compute the local current density $j$ that depends on the voltage at the corresponding point and use it to compute the overall current for the cell. Thus, we have
\begin{equation}
   j = j_{L} - j_{0}\cdot (\text{exp}(qV/k_{\text{B}}T) - 1).
\end{equation}

The overall current flow on the solar cell front surface can be modeled using the physics of the conductive layer on the front surface. It can be explained using the reformulated version of Ohm's law \cite{darrigol2000electrodynamics}, stated as
\begin{equation}
    \mathbf{j} = \sigma \mathbf{E},
    \label{eq_cellphy}
\end{equation}
where $\mathbf{j}$, $\sigma$ and $\mathbf{E}$ denote current density, material dependent conductivity and the electric field in a given area. For a conservative field, using Kirchoff's circuit law and discretizing the physics using finite element model, the relation between $I$ and $V$ can be stated in terms of matrix equation as

\begin{equation}
    \mathbf{G}\mathbf{V} = \mathbf{I},
\end{equation}
where \textbf{G} is the total conductivity matrix, \textbf{V} is the voltage vector and \textbf{I} is the current vector. We point out here that $\mathbf{G} = \mathbf{G(\mathbf{x})}$ is a function of the metallization design $\mathbf{x}$ itself. See \cite{Gupta2015smo} for the mathematical details. 

The output power from the solar cell, computed at the busbar location, can then be written as
\begin{equation}
   P_{\text{out}} = V_{\text{bus}}\sum \mathbf{I},  
   \label{eq_power}
\end{equation}
where $V_{\text{bus}}$ is the applied busbar voltage and $\sum \mathbf{I}$ denotes the sum of all elements of matrix \textbf{I}. The different configurations have been tested with $V_{bus}$ =  0.5. Efficiency ($\eta$) is calculated as
\begin{equation}
    \eta = \frac{P_{\text{out}}/A_c}{P_{\text{in}}} \times 100%
\end{equation}
where $P_{in}$ is the input power for the cell and $A_c$ denotes cell area. The standard value is \mbox{1000 Wm$^{-2}$}. $P_{\text{out}}$ is the power produced by the cell.

\section{Model Parameterization with CNN}
\label{others}
It has already been shown in previous works that convolutional neural networks (CNN) exhibit inherent bias for image processing tasks such as denoising, among others. Inspired from the work of \cite{Hoyer}, we show that this holds true for highly-nonlinear physics-based optimization tasks as well. For this, we convert the domain of traditional density-based TO to that of learning weights of a CNN. This is achieved through reparameterization of the design space of TO.  Reparameterization (also called change of variable), in simple terms, is the change of basis of a function from one domain to another. In the traditional TO-based designing of the metallization patterns, the density variables are optimized such that the output power of the solar cell is \mbox{maximized \citep{Gupta2015smo}}.

\begin{figure*}
    \centering
	  \includegraphics[scale=0.45]{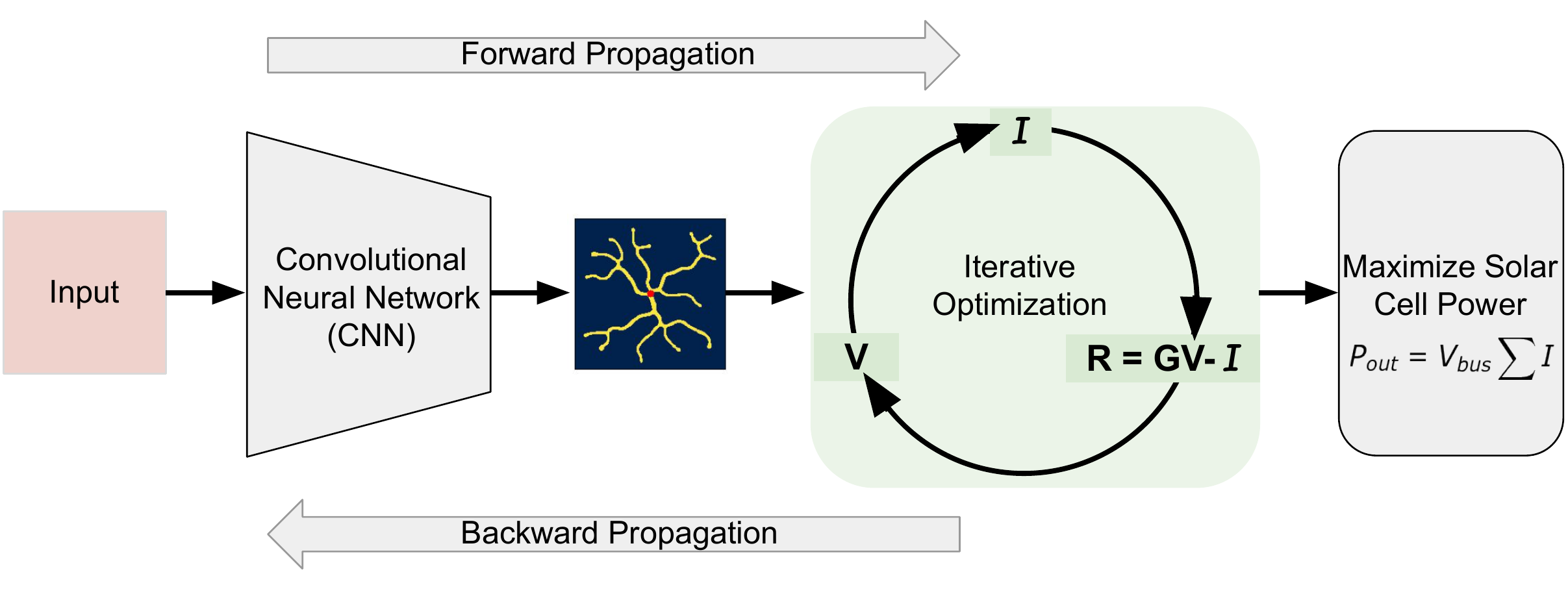}
	   \caption{Schematic diagram showing the workflow of SolarNet. The weights of the CNN are optimized and the density field generated by the CNN is validated using the physics of the problem. The model is trained end-to-end using power loss and error is backpropagated through the mathematical modelling steps as well as the CNN pipeline.}
	\label{fig-schematic}
\end{figure*}

Rather than solving the problem outlined above using a purely data-driven approach, we transform this problem into that of optimizing the weights of a CNN such that the density values are produced as its learnt output. Schematic representation of the workflow is shown in Fig. \ref{fig-schematic}.  In simple words, the CNN model acts like an estimator that returns an image representation of a solar cell depending upon the weights, biases, input data and activation functions. Each pixel in the image lies in the range [0,1] and corresponds to density value at that point. The generated density values are further used to calculate the power output of the solar cell using the original mathematical model itself. The entire optimization involving training of the CNN weights as well as well as getting physically correct TO designs is performed end-to-end. The gradient for the loss function backpropagates through the mathematical model of the solar cell followed by through the CNN pipeline.  

For the backpropagation of error through the mathematical model, we employ an adjoint sensitivity analysis scheme \cite{Gregoire}. Adjoint sensitivity analysis approach allows faster computation of the gradients with respect to all the design parameters. Parts of the gradients are directly computed using automatic differentiation module in AutoGrad (\cite{autograd}). Details on computing the derivatives can be found in \cite{Gupta2015smo}.
\section{Experiments}

\textbf{Experimental setup. }We demonstrate the efficacy of SolarNet, the proposed reparameterization strategy for solar cells, on two different shapes: \emph{square} and \emph{triangular}. Further, we also investigate the effect on solar cells with 4 different busbar configurations. All experiments are modeled with thin-film solar cells, and details related to the various cell parameters are described in Table \ref{table-cell-params}. For the TO method, radius of density filtering (r) is set to 1.5 elements. 

\begin{table}
\vspace{-1em}
\centering
\caption{Solar cell parameters}
\begin{tabular}{l | l}
 \toprule
 Parameter & Value \\
 \midrule
   Cell Length ($L_{y}$) & 1.5 cm \\
   Cell Width ($L_{x}$) & 1.5 cm  \\
   Busbar width & 2 mm \\
   Conductivity & 100 S$^{-1}$ 
   \\ 
   Minimum Conductivity & 0.02 S$^{-1}$
   \\
   Photo-illumination current - $j_{l}$ & 310 Am$^{-2}$
   \\
   Reverse Bias Current - $j_{0}$ & -0.06 Am$^{-2}$
   \\
   Busbar voltage $V_{\text{bus}}$ & 0.5 V
   \\
   Temperature $T$ & 320 K
   \\
   Mean illumination intensity & 1000 Wm$^{-2}$ \\
   
 \bottomrule
 \end{tabular}
 \label{table-cell-params}
\end{table}

For the CNN model, we use an architecture composed of 5 convolutional layers, each of kernel size $5 \times 5$ and followed by a batch normalization and a Leaky ReLU layer, except the last layer that uses a sigmoid function. The CNN model is written in Tensorflow 2.0, and derivatives for the mathematical model of electrical conduction are obtained using the Autograd package. The weights of the CNN are initialized using Glorot normal initializer and an output of dimensions $200 \times 200$ is generated for all cases.

\begin{figure*}
    \centering 

\begin{subfigure}{0.2\textwidth}
  \includegraphics[width=\linewidth]{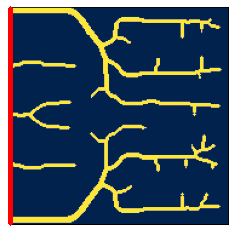}
  \caption*{12.20\%}
\end{subfigure}
\begin{subfigure}{0.2\textwidth}
  \includegraphics[width=\linewidth]{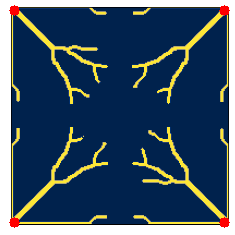}
  \caption*{12.05\%}
\end{subfigure}
\begin{subfigure}{0.2\textwidth}
  \includegraphics[width=\linewidth]{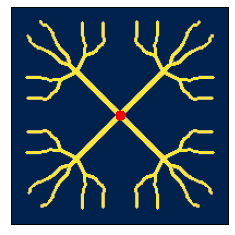}
  \caption*{11.98\%}
\end{subfigure}
\begin{subfigure}{0.2\textwidth}
  \includegraphics[width=\linewidth]{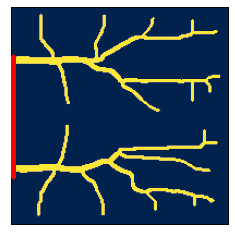}
  \caption*{12.55\%}
  
\end{subfigure}

\medskip
\begin{subfigure}{0.2\textwidth}
  \includegraphics[width=\linewidth]{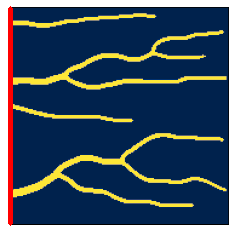}
  \caption*{12.34\%}
  
\end{subfigure} 
\begin{subfigure}{0.2\textwidth}
  \includegraphics[width=\linewidth]{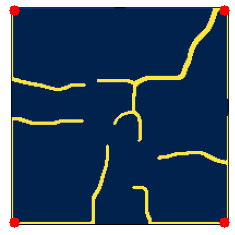}
  \caption*{12.47\%}
  
\end{subfigure}
\begin{subfigure}{0.2\textwidth}
  \includegraphics[width=\linewidth]{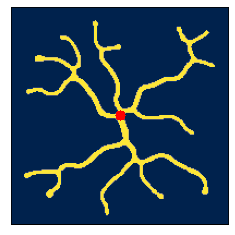}
  \caption*{12.25\%}
  
\end{subfigure}
\begin{subfigure}{0.2\textwidth}
  \includegraphics[width=\linewidth]{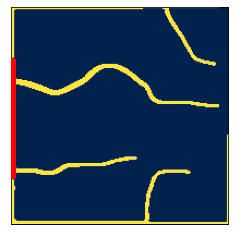}
  \caption*{12.64\%}
\end{subfigure}
\caption{Optimized metallization schemes obtained using the traditional density-based TO (top) and SolarNet reparameterization (bottom) for 4 different busbar configurations. Busbar locations are marked in red.}
\label{fig-square-cells}
\vspace{-1em}
\end{figure*}

\begin{figure}
\vspace{-1em}
\centering
\begin{subfigure}{0.19\textwidth}
\centering
  \includegraphics[width=25mm, height=30mm]{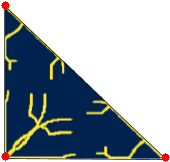}
  \caption{12.31\%}
\end{subfigure} \hspace{1em}
\begin{subfigure}{0.19\textwidth}
\centering
  \includegraphics[width=25mm, height=30mm]{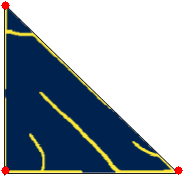}
  \caption{12.67\%}
\end{subfigure}
\caption{Optimized metallization schemes obtained using the traditional density-based TO (left) and SolarNet reparameterization (right) for triangular solar cell domain. Busbar locations are marked in red.}
\label{fig-triangle-cells}
\vspace{-1em}
\end{figure}

\textbf{Results. }Fig. \ref{fig-square-cells} shows the results obtained for the traditional density-based TO scheme as well as our SolarNet reparameterization. Interestingly, SolarNet outperforms the performance scores of the traditional TO approach for all the busbar configurations of the square-shaped cells. The improvements go up to 0.42\% in absolute terms, which amounts to a relative improvement in 3.5\% in the efficiency of the solar cell. We believe that this improvement is primarily because SolarNet can exploit locally optimal asymmetric designs even for symmetric boundary conditions. However, the traditional TO method is only able to produce symmetric designs. 

SolarNet also improves performance of solar cells on freeform non-conventional geometries, and this is shown in Fig. \ref{fig-triangle-cells}. For the triangular design domain, SolarNet outperforms the traditional approach by some extent. We believe that the improvements with SolarNet observed across different geometries and busbar conditions are primarily because SolarNet can exploit locally optimal asymmetric designs even for symmetric boundary conditions. However, the traditional TO method is only able to produce symmetric designs.

\textbf{Discussion. } We demonstrated improvements of up to 3.5\% in efficiency of the solar cells in relative terms.  
However, in absolute terms, it is less than 0.5\% for all the cases reported in the paper. We point out that this small absolute improvement is still significant. At the standard conditions of AM 1.5, where one sun intensity is assumed to be 1000 Wm$^{-2}$, an improvement of 0.1\% in efficiency implies an increase of 1 W in power for an  irradiance area of 1 m$^2$.
\section{Conclusions}

This paper presents SolarNet, a CNN-based reparameterization strategy for the optimization of metallization designs in solar cells. SolarNet facilitates optimizing directly the weights of CNN and generating the topological design of the front metallization as its output. Through experiments on multiple solar cells shapes and busbar geometries, we have shown that SolarNet improves the performance of solar cells over the standard topology optimization scheme. Clearly, the results indicate that deep learning strategies such as CNN exhibit potential for physics-based optimization problems, and further research in this direction is of interest.

\bibliography{iclr2021_conference}
\bibliographystyle{iclr2021_conference}


\end{document}